\documentclass[a4paper,11pt]{article}
\pdfoutput=1
\usepackage{jcappub}
\usepackage[T1]{fontenc}
\usepackage{amsfonts,amsmath,amssymb}
\usepackage{graphicx}
\usepackage{color}
\usepackage{natbib}
\title{Search for Lorentz Invariance Violation from stacked Gamma-Ray Burst  spectral lag data}
\author[a]{Rajdeep Agrawal,}
\author[a]{Haveesh Singirikonda,}
\author[a,1]{and Shantanu Desai
\note{Corresponding author.}}
\affiliation[a] {Department of Physics, Indian Institute of Technology, Hyderabad, Telangana-502285, India}

\emailAdd{ep18btech11012@iith.ac.in}
\emailAdd{ep17btech11010@iith.ac.in
}
\emailAdd{shantanud@phy.iith.ac.in}

\abstract{A number of works have claimed detections of a turn-over in the spectral lag data for individual Gamma-Ray Bursts (GRBs),  caused by an energy-dependent speed of light, which could be  a possible manifestation of Lorentz invariance violation (LIV). Here, we stack the spectral lag data from a total of 37 GRBs (with  a  total of 91 measurements), to verify if  the combined data  is consistent  with  a unified model consisting of intrinsic astrophysical emission, along with another contribution due to LIV.  We then carry out Bayesian model comparison to ascertain if this combined spectral lag data shows a preference for an energy-dependent speed of light, as compared to only an  intrinsic astrophysical emission mechanism. We do not find a decisive evidence for such an energy-dependent speed of light for two different models of LIV. When we assume a constant intrinsic lag coupled with an unknown intrinsic scatter, we do not find any evidence for LIV. However, when we use  GRB-dependent parameters to model the intrinsic emission, we get decisive evidence for LIV violation. We then carry out a search for LIV Standard Model Extension using this dataset as well as an independent search using a separate dataset consisting of rest-frame spectral lags.   Finally,  none of the   models considered  here with any of the aforementioned assumptions provide a good fit to the stacked spectral lag data,  indicating that there is still missing Physics in the model for intrinsic spectral lags.}
\begin{document}

\maketitle
\flushbottom
\section{Introduction}

In special relativity, the speed of light, $c$, is a Lorentz invariant quantity. However, this 
\emph{ansatz} is not valid in many theories beyond the Standard Model of Particle Physics and also various quantum gravity and string theory models (See ~\cite{Tasson,Mattingly,GAA,Wei2021} for recent reviews). In these models, Lorentz invariance is expected to be broken at very high energies close to the Planck scale ($E_{pl} \sim 10^{19}$ GeV), and
the speed of light is a function of the  energy of the associated photon~\cite{GAA98}. Alternately,   one can think of  the vacuum refractive index as been  different from unity in such models~\cite{Ellis}. The energy-dependent speed of light can then  be written as~\cite{AmelinoCamelia98}:
\begin{equation}
    v(E) = c\left[1 - s_{\pm}\frac{n+1}{2} \left(\frac{E}{E_{QG}}\right)^n\right],
    \label{eq:vE}
\end{equation}
where $s_{\pm} = \pm 1$ denotes the sign of the Lorentz Invariance violation (LIV), corresponding to sub-luminal ($s_{\pm}=+1$) or super-luminal ($s_{\pm}=-1$); $E_{QG}$ denotes the quantum gravity scale where LIV effects kick in, and $n$ is a model-dependent term and is usually equal to one or two, corresponding to linear or quadratic LIV.  The values of $n$ for different LIV models can be found in ~\cite{Pan}.

A plethora of searches for LIV  have been carried out using photons, neutrinos, and gravitational waves, by looking for the energy dependent speed of light as given by  Eq.~\ref{eq:vE}, with $s_{\pm}=+1$. The astrophysical sources used for searches of LIV with photons include pulsars~\cite{Kaaret,Magiccrab}, AGNs~\cite{AGN,Ellis13,HESS,Friedman}, and Gamma-Ray Bursts (GRBs, hereafter) ~\cite{AmelinoCamelia98,Ellis03,Ellis,Abdo,Chang,Vasileiou13,Vasileiou15,Zhang,Liu,Pan15,Xu1,Xu2,Wei17,Ellis19,Weipolarization,Du,Pan,MAGIC,Zou}. Reviews of all these astrophysical searches for LIV with these sources can be found in ~\cite{Ellis13,Bolmont,Horns}.
The corresponding results on LIV with neutrinos can be found in~\cite{AC2016,AC2,Huang,Huang2,kaiwang,Ellisneutrino}. A constraint on LIV using the first gravitational wave event GW150914 has also   been obtained~\cite{EllisGW}.

The observable used in almost all the studies of LIV with GRBs consists of spectral lags, which can be  defined as the difference in the  arrival times  between high energy and low energy photons, and is positive if the high energy photons arrive earlier than the low energy ones. 
The first such systematic study with a large GRB sample was carried out by ~\cite{Ellis}, who considered a sample of 35 GRBs in the redshift range $z=0.168-4.3$ from HETE, BATSE, and Neil Gehrels SWIFT. The spectral lags were modeled  as the sum of a constant intrinsic lag  together with another contribution due to   an energy-dependent speed of light. They also  found a $4\sigma$  evidence  for the  higher energy photons to arrive earlier than the lower energy ones, which at face value points to evidence for LIV~\cite{Ellis}. However, when an additional systematic offset was added to make the $\chi^2$/DOF equal to one,  the statistical significance for LIV reduced to about $1\sigma$. Subsequently, they set a lower limit of $E_{QG} \geq (0.9-2.1) \times 10^{16}$ GeV at 95\% c.l.  

The first convincing case for a spectral lag turnover from positive to negative lags, using multiple spectral lag data from a single GRB (GRB 160625B) was made 
in ~\cite{Wei}. This work modeled the time-lag data as a sum of intrinsic astrophysical time-lag  and another  lag due to the  energy-dependent speed of light from LIV. All previous works prior to ~\cite{Wei} had assumed a constant intrinsic lag in the source frame. The intrinsic time delay proposed in ~\cite{Wei}  was a  phenomenological model,  parameterized as function of the energy.
 They found that the spectral lag for this  GRB  shows a turnover at around 8 MeV, indicating a transition from positive to negative lags. They argued that this transition could be a signature of LIV, which kicks in at high energies. Their best-fit value for  $E_{QG}$    was $\log(E_{QG}/GeV)$  are $15.66^{+0.55}_{-0.01}$ and $7.17^{+0.17}_{-0.02}$ for linear and quadratic LIV models, respectively~\cite{Wei}.  The statistical significance for this spectral lag transition corresponds to a $Z$-score 
between $3.05-3.74\sigma$ (using frequentist techniques),  and $\Delta$AIC/BIC $>10$ for the quadratic LIV model~\cite{Ganguly}. The information theory techniques therefore point to decisive evidence for the quadratic LIV model.

Subsequently, ~\cite{Pan} stacked the spectral lag data for GRB 160625B along with the data for 35 GRBs from ~\cite{Ellis} and argued that this provides a more robust estimate of the intrinsic time lag.  Their best-fit estimates for  $\log(E_{QG}/GeV)$ are $14.523^{+0.022}_{-0.025}$ and $8.79 \pm 0.0097$
for linear and quadratic LIV models, respectively.   Most recently, a similar spectral lag transition from positive to negative lags (similar to GRB 1606025B)  was detected in GRB 1901114C at $\sim$ 0.7 MeV~\cite{Du}. However, the model for the intrinsic time lag asserted  was opposite in sign compared to ~\cite{Wei,Pan}. The statistical significance of the spectral lag transition was also estimated using Bayesian model comparison and found to be $>100$, pointing to decisive evidence using Jeffrey's scale. Similar to ~\cite{Wei,Pan}, they obtained best-fit values for linear and quadratic models, with  their 2$\sigma$ bounds  given by $\log(E_{QG}/GeV)$= $14.49^{+0.12}_{-0.13}$ and $6.00 \pm 0.06$, respectively. Furthermore, they also constrained the parameters of LIV Standard Model Extension models~\cite{Kostelecky}. They also computed the $\chi^2/DOF$ for the null hypothesis (consisting of only intrinsic emission) as well as all the LIV models considered. They showed that all the LIV models have $\chi^2$/DOF are less than or close to one, indicating a good fit to the LIV hypothesis~\cite{Du}. The MAGIC collaboration however failed to find a similar evidence for an energy-dependent speed of light in the TeV gamma ray data for the same GRB. Using conservative assumptions on spectral and temporal evolution, the MAGIC collaboration  set a lower limit of $\mathcal{O} (10^{19})$ and $\mathcal{O}(10^{10})$ GeV for linear and quadratic models, respectively~\cite{MAGIC}.

We note that even though some of  the above aformentioned analyses using spectral lag data, obtained bound $1\sigma$ confidence intervals for $E_{QG}$, which is less than the Planck energy scale, the conclusions of these works only reported one-sided lower limits on $E_{QG}$, which is same as the central estimate~\cite{Wei,Pan,Du}. This does not adhere to the formal way of calculating one-sided lower (or upper) limits recommended by the PDG~\cite{PDG}.

Furthermore, as pointed out in ~\cite{Ganguly}, the estimated LIV energy scale estimated in ~\cite{Wei} would contradict previous limits by approximately  3-4 orders of magnitude~\cite{Abdo,Vasileiou13}. This point would also apply to the recent results in ~\cite{Pan,Du}. Still, given the tantalizing hints for LIV in each  of the aforementiond works~\cite{Ellis,Wei,Pan,Du}, if the spectral lag data for all the GRBs can be described by a unified model, one would expect the net statistical significance of the LIV  to get  enhanced, if we stack the spectral lag data from  all the GRBs, and analyze them uniformly with the same model. In this work, we therefore stack the  data from all the three  works which found these hints for LIV signatures in the spectral lag data~\cite{Ellis,Pan,Du}. We then do a Bayesian model comparison of the hypothesis that the combined spectral lag data is a combination of both an intrinsic and a  LIV-induced lag, as compared to only an intrinsic lag due to astrophysical emission. The model assumed for the intrinsic astrophysical emission is same as that proposed in ~\cite{Wei}, which  has also been used in ~\cite{Pan}. 

The outline of this paper is as follows. We briefly discuss the model comparison technique in Sec.~\ref{sec:modelcomp}.  The datasets used for this analysis are described in Sec.~\ref{sec:datasets}. The analysis procedure used to analyze the GRB spectral lag data is discussed in Sec.~\ref{sec:analysis}. The reconstruction of the cosmic expansion history using chronometers is described in Sec.~\ref{sec:mi}. Our main results are presented in Sec.~\ref{sec:results}. A variant of this analysis using a constant intrinsic lag is presented in Sec.~\ref{sec:const}, and with different intrinsic parameters in Sec.~\ref{sec:diff}. A search for LIV using rest-frame spectral lags is outlined in Sec.~\ref{sec:sourceframe}. Constraints on LIV SME parameters can be found in Sec.~\ref{sec:SME}. We conclude in Sec.~\ref{sec:conclusions}.

\section{Bayesian Model Comparison}
\label{sec:modelcomp}
Since a central theme of this work is model selection, we give a brief primer to these techniques. There are three main methods used for model comparison: frequentist, information theory, and Bayesian. A comparison and contrast of these methods can be found in ~\cite{Liddle,Liddle07,Shi,Sanjib,Weller,Trotta,Gelman,astroml}. In our previous works, we have applied all these techniques to a large number of model selection problems in Astrophysics and Cosmology~\cite{Ganguly,Desai16a,Desai16b,Kulkarni,Kulkarniexoplanet,Haveesh,Krishak1,Krishak2,Krishak3,Krishak4}. In this work, we shall only use the Bayesian method for model comparison, as this is the most robust among the various techniques and does not involve any assumptions~\cite{Sanjib}. We provide a brief prelude to the Bayesian model comparison technique. For more details, the reader can refer to ~\cite{astroml,Weller,Sanjib,Trotta} (and references therein).

Bayesian Model comparison is based on the Bayes Theorem in probability, which states: \begin{equation}
        P(M|D) = \frac{P(D|M)P(M)}{P(D)}
        \label{eq:bayesthm}
    \end{equation} 
for a model $M$ with respect to data $D$. Here, $P(M|D)$ represents the posterior probability and $P(D|M)$ is the marginal likelihood, also known as the Bayesian Evidence. This can be defined as:
\begin{equation}
P(D|M) = E(M) = \int P(D|M, \theta)P(\theta|M) \, d\theta
\label{eq:evid}
\end{equation}
where $\theta$ is the vector of parameters associated with the model $M$, $P(\theta|M)$ is  the prior on the the parameter vector ($\theta$) for that model. 
To perform model comparison between two models $M_1$ and $M_2$, we calculate the Posterior odds ratio, which is the ratio of their posterior probabilities. So, the odds ratio for model $M_2$ over $M_1$ is given by: 
\begin{equation}
O_{21} = \frac{P(M_2|D)}{P(M_1|D)}
\end{equation}
Using Eq.~\ref{eq:bayesthm} and ~\ref{eq:evid}, this can be further written as: \begin{equation} 
    O_{21} = \frac{E(M_2)P(M_2)}{E(M_1)P(M_1)} = B_{21}\frac{P(M_2)}{P(M_1)}\end{equation}
where the term $B_{21}$ is the Bayes Factor, given by the ratio of Bayesian Evidences of the two models. If we were to assume equal apriori probabilities for both the models, the Odds Ratio is the same as Bayes Factor. Therefore, we obtain:
\begin{equation} O_{21} = B_{21} = \frac{\int P(D|M_2, \theta_2)P(\theta_2|M_2) \, d\theta_2}{\int P(D|M_1, \theta_1)P(\theta_1|M_1) \, d\theta_1} \end{equation}
The Bayes factor is then used for Bayesian model comparison.  Note that unlike other model selection techniques, Bayesian model comparison does not involve the computation of best-fit parameters for a given model.

The model with the larger value of Bayesian evidence will be considered as the favored model. We then use the Jeffrey's scale to qualitatively  assess the significance of the favored model~\cite{Trotta}. According to this scale, a Bayes Factor $< 1$ indicates negative support for the model in the numerator ($M_2$), thereby favouring the model $M_1$. A value exceeding 10 implies strong evidence for $M_2$, while a value  greater than  100 indicates decisive evidence.  Therefore, a smoking gun evidence for model $M_2$ over $M_1$ requires a Bayes factor greater than 100.

\section{Datasets used for the analysis}
\label{sec:datasets}
We briefly describe each of the datasets analyzed in ~\cite{Ellis,Wei,Du} that are used for our stacked analysis. More details can be found in the original papers and references therein. 

Ellis et al \cite{Ellis} considered a sample of 35 different GRBs (with one spectral lag per GRB) in the redshift range $0.25<z<6.29$, detected by three different telescopes, viz. BATSE, HETE, and Neil Gehrels SWIFT.
For BATSE  data, the spectral lags were measured  in the 115-320 keV band relative to the 25-55 keV band.  The Neil Gehrels SWIFT  and HETE data were also normalized   to the same energy bands. The spectral time-lag data  along with the errors for all the 35 GRBs can be found in Table 1 of ~\cite{Ellis}. The lower and higher energy intervals corresponding to these time lags were assumed to be the centers of the energy bins, viz. 40 keV and  217.5 keV, respectively. The error in energy difference was calculated from the quadrature sum of the half-widths for the two energy bins.

The second dataset used in this work is the spectral lag data for GRB 160625B analyzed in \cite{Wei}.  This GRB is located at a redshift of $z=1.41$.  This GRB was detected by both Fermi-GBM and Fermi-LAT. The light curve for this GRB contained three isolated sub-bursts, lasting about 770 seconds. This long duration facilitated the measurement of  37 independent spectral lags in the 15-350 keV intervals, with respect to a fixed  energy of 11.34 keV. The spectral lag data along with their error bars can be found in Table 1 of ~\cite{Wei}. The error in the energy difference was equal to the half-width of each energy bin.

The final dataset we used consists of 19 spectral lag measurements of GRB190114C, located at a redshift of $z=0.4245$~\cite{Du}. This GRB was detected by the Neil Gehrels SWIFT telescope with $T_{90}$ equal to 362 and 116 seconds in the 15-350 keV and 50-300 keV energy bands, respectively. This GRB was also detected by Fermi-LAT at MeV energies and by the MAGIC telescope up to   TeV energies~\cite{MagicGRB}.  Using the Fermi-GBM light curves, ~\cite{Du} constructed the spectral lag  data from  15 keV to 5000 keV, compared to the lowest energy value of 12.5 keV. These spectral lags along with their errors can be found in Table 1 of ~\cite{Du}. Similar to GRB 160625B, we used the half-widths of each energy bin to characterize the  error for energy value.
 
Finally, we combine all the aforementioned spectral lag measurements. We therefore get a total of 91 spectral lag measurements from 37 GRBs, located in a redshift range from 0.25 to 6.29. Since all the GRB redshifts were  spectroscopic, we do not consider the error in their redshifts for our analysis, as they are expected to be negligible. 

\section{Model used for time lags}
\label{sec:analysis}
The observed spectral time lag ($\Delta t_{obs}$) for a given GRB, corresponding to an energy interval $\Delta E$ can be modeled as a sum of two independent delays: 
\begin{equation} 
\Delta t_{obs} = \Delta t_{int}^{obs} + \Delta t_{LIV} 
\label{eq:delta t}
\end{equation}
where $\Delta t_{obs}$ is the observed spectral lag;   $\Delta t_{int}^{obs}$ is the intrinsic time delay due to astrophysical emission in the observer frame, and $\Delta t_{LIV}$ due to LIV. 

Many initial works in the searches for LIV, assumed a constant intrinsic time delay in the source frame~\cite{Ellis,Shao,Pan15,Xu1,Xu2,Liu,Wei17,Zou}. We shall use the following model for the intrinsic time emission in the source frame~\cite{Wei},
\begin{equation} 
\Delta t_{int} =\tau\Big[ \Big(\frac{E}{keV}\Big)^{\alpha}-\Big(\frac{E_0}{keV}\Big)^{\alpha}\Big]
\label{eq:delt_int}
\end{equation}
This model for the intrinsic emission in the source frame   was obtained from analyzing the statistical properties of 50 single-pulsed GRBs~\cite{Shao}.  This parametric form  has been used for LIV searches using spectral lags in a number of works~\cite{Wei,Wei2,Wei17,Ganguly,Pan}. Du et al~\cite{Du} used a slight variant of Eq.~\ref{eq:delt_int}, where  $\alpha$ was replaced by $-\alpha$ and $\tau$ by -$\tau$. For this analysis, we shall choose broad priors on $\alpha$ and $\tau$ to encompass both these variants. We note however that  the detailed emission mechanism of GRBs is still unknown, and there could be other astrophysical parameters characterizing the intrinsic mechanism. Ref.~\cite{Du19} 
has found a correlation between spectral and spectral evolution. We shall explore other models for intrinsic emission and their impact on LIV searches in the forthcoming sections. In this work however, use Eq.~\ref{eq:delt_int}, since it has also been used in the earlier works.

The intrinsic time delay in the observer frame is given by 
\begin{equation}
 \Delta t_{int}^{obs} =(1+z)  \Delta t_{int},
 \label{eq:nullhypothesis}
\end{equation}
where the $(1+z)$ term accounts for the cosmological time dilation. For the null hypothesis, the spectral  lags will be directly fitted to only  Eq.~\ref{eq:nullhypothesis}.

The time delay due to linear and quadratic LIV models can be obtained from $n=1$ and $n=2$, respectively of the following equation~\cite{Jacob},
\begin{equation}
\Delta t_{LIV} = t_0-t  = -\left(\frac{1+n}{2H_0}\right)\left(\frac{E^n - E_0^n}{E_{QG,n}^n}\right)\int_{0}^{z} \frac{(1+z^{\prime})^n}{h(z^{\prime})} \, dz^{\prime} 
\label{eq:deltaliv}
\end{equation} 
where $E_{QG,n}$ is the Lorentz-violating or quantum gravity scale, above which Lorentz violation is turned on. In Eq.~\ref{eq:deltaliv}, $n=1$ and  $n=2$ denote  linear  and  quadratic LIV models, respectively. Finally,
$h(z) \equiv \frac{H(z)}{H_0}$ is the dimensionless  Hubble parameter as a function of redshift. For the current standard $\Lambda$CDM model~\cite{Planck18}, $h(z)= \sqrt{\Omega_M (1+z^\prime)^3 + \Omega_\Lambda}$ and this parametric form has been used for the analysis in ~\cite{Ellis,Wei,Ganguly,Wei17,Du}.  Here, we reconstruct $h(z)$ in a non-parametric manner using Gaussian Process Regression (GPR) similar to the analysis in ~\cite{Pan}. The integral in Eq.~\ref{eq:deltaliv} can be  encompassed in the parameter $K(z)$ defined as 
\begin{equation}
K(z) = \int_{0}^{z} dz' \frac{(1+z^{\prime})^n}{h(z^{\prime})}
\label{eq:Kz}
\end{equation}
The estimation of $K(z)$ in a model-independent fashion using GPR  will be discussed in Sec.~\ref{sec:mi}.

The last ingredient which we need for our analysis is the total error ($\sigma_{tot}$) in the observables. This includes the error in  the observed spectral  lag ($\sigma_t$) as well as the error in the    energy ($\sigma_E$). We use the prescription in ~\cite{Weiner} to combine these errors and the total error $\sigma_{tot}$ is given as:
\begin{equation} 
\sigma_{tot}^2 = \sigma_t^2 + \Big(\frac{\partial f}{\partial E}\Big)^2\sigma_E^2 
\label{eq:totalerror}.
\end{equation}
where $f$ corresponds to  the particular model  been tested.

\section{Reconstruction of Expansion History}
\label{sec:mi}
In this section, we shall discuss the calculation of Eq.~\ref{eq:Kz} in a model-independent fashion using cosmic chronometers, which is agnostic to any particular Cosmology.
\subsection{Cosmic Chronometers} 
\label{sec:CCdata}

In Equation \ref{eq:Kz}, $K(z)$ is generally obtained from the underlying cosmological model. Using GPR, we obviate this  requirement and reconstruct  $h(z)$ directly from the  data. We use Hubble parameter measurements   from cosmic chronometers (CC) \cite{Jimenez_2002}, which is a model independent method for  measuring the Hubble parameter using the redshifts and relative ages of galaxies.  For this work, we use the use the same chronometer dataset, as that in ~\cite{Haveesh}, which consists of  31 data points covering the redshift range $0.07 < z < 1.965$~\cite{Li}. 

The cosmic chronometer technique uses the following way of  defining the Hubble parameter~\cite{Jimenez_2002}:
\begin{equation}
    H(z) = - \frac{1}{1+z} \frac{dz}{dt}
\end{equation}
For small changes in $z$ and $t$, this relation can be generalized to 
\begin{equation}
    H(z) = - \frac{1}{1+z} \frac{\Delta z}{\Delta t}
\end{equation}
Therefore, through the measurements of redshifts and ages from spectroscopic analysis of galaxies in this formula, the value of the Hubble parameter ($H(z)$) can be estimated at a particular redshift $z$. The added advantage here is that only the relative ages of the galaxies are needed for this measurement.  
Once we obtain sufficient measurements of $H(z)$ at different redshifts, one can interpolate (or extrapolate) between these measurements to get $H(z)$ at any input redshift. Therefore,  no underlying cosmological model is used to reconstruct $H(z)$. We now discuss the GPR technique used for reconstructing the expansion history at any redshift using these chronometers.

\subsection{Estimation of $h(z)$ using Gaussian Process Regression}
\label{sec:gpr}

The GPR technique  has seen increased usage in Astronomy literature as it has proved to be a useful tool for model-independent analyses (see ~\cite{Haveesh} and references therein). Simply put, the purpose of GPR is to reconstruct a quantity from a data set without the assumption of a parametric model. The popular choices for model-independent analyses are to assume a parametrization, such as different kinds of polynomials, for a function. This choice of parameterization can be arbitrary sometimes and there is  no rule set in stone about which choice is better. The advantage of using GPR is that it completely removes the need for choosing a parametric form for a given function, and can still reconstruct the quantity at all values.

The main premise behind this technique lies in the idea of Gaussian distribution extended to functions. One requirement for this method is that the errors in the data used must be Gaussian. A covariance function is used to connect the points at which the data is available to the other points in space. This covariance function will help in predicting the value at this point from the information available from the data set. The most common choice for the covariance function is the squared exponential function, which is given by
\begin{equation}
	k(x,\tilde{x}) = \sigma^2 exp\left( - \frac{(x-\tilde{x})^2}{2l^2} \right) 
\end{equation}

This function is the simplest choice which can serve the purpose for GPR. There are many other alternatives like the Mat\'ern and Cauchy kernels. This covariance function can be written in the form of a matrix, for a set of input points $\mathbf{X}$ as 
\begin{equation}
	[\mathbf{K}(\mathbf{X},\mathbf{X})]_{i,j} = k(x_i,x_j)
\end{equation}

Another requirement for this is the choice for the function $\mu(x)$, which is the  apriori mean of this quantity. A constant function is a good choice for this. Now this can be used to extrapolate and determine the mean and errors at other points in space. 

\begin{eqnarray}
 \langle \mathbf{f^*} \rangle &=& \boldsymbol\mu^* + \mathbf{K}(X^*,X) [\mathbf{K}(X,X)+\mathbf{C} ]^{-1}(\mathbf{y-\mu}), \nonumber\\
cov(\mathbf{f^*}) &=& \mathbf{K}(X^*,X^*)  - \mathbf{K}(X^*,X)[\mathbf{K}(X,X)+\mathbf{C}]^{-1} \mathbf{K}(X,X^*),
\label{eq:mean_cov_f}
\end{eqnarray}
where $\mathbf{X^*}$ represents the points at which we want to predict the values for the quantity $f(x)$, $\langle \mathbf{f^*} \rangle$ is the mean value predicted for the function $f(x)$ at $\mathbf{X^*}$, $cov(\mathbf{f^*})$ is the error on these values, $\mathbf{y}$ is the set of values of $f(x)$ available from the data, and $\mathbf{C}$ is the covariance matrix for the data set (for uncorrelated errors, this is simply $diag(\sigma_i^2)$). The set $\mathbf{X}$ represents the data set. A much more detailed explanation of GPR can be found in \cite{Seikel_2012}. We implement GPR using the publicly available code {\tt GaPP} in Python \cite{Seikel_2012}.

After a successful reconstruction  of $H(z)$  (and $h(z)$) using GPR, we can estimate  the value of  $h(z)$ at any redshift along with its $1\sigma$ error bars.   Using this reconstructed $h(z)$, the integral in Eq.~\ref{eq:Kz}, required to compute $K(z)$ can be computed using any standard numerical integration algorithm. In this work, we use the {\tt quad} function in the  {\tt scipy} Python module.

\section {Analysis and Results}
\label{sec:results}
We now discuss the analysis of the stacked spectral lag data. The first step in any model comparison involves parameter estimation. For this purpose, we  write down the following likelihood ($\mathcal{L}$) for the given model:
\begin{equation}
     \mathcal{L}=\prod_{i=1}^N \frac{1}{\sigma_t \sqrt{2\pi}} \exp \left\{-\frac{[\Delta t _i-f(\Delta E_i,\theta)]^2}{2\sigma_t^2}\right\},
     \label{eq:likelihood}
  \end{equation}
where $\Delta t_i$ denote the spectral lag data corresponding to the energy intervals $\Delta E_i$; $\sigma_t$ denotes the total uncertainty  as discussed in Eq.~\ref{eq:totalerror}; $f(\Delta E_i,\theta)$ is the hypothesis used to fit the data (cf. Eq.~\ref{eq:delta t});
and $\theta$ is the vector of parameters used to fit each hypothesis. For the null hypothesis, $\Delta t_{LIV}$  would be equal to 0. The best-fit values for each of the models are obtained by maximizing the posterior $P(\theta|D,M) \propto P(D|M,\theta) P(\theta)$~\cite{Trotta}, where $P(\theta)$ represents the priors for each of the models. The priors used for each of the three models can be found in Table~\ref{priortable}.

Although our main goal is model comparison, we would like to get  a sense of how good the best fits for each of the models are. For that, we use $\chi^2 = -2 \ln{\mathcal{L}}$, where  $\mathcal{L}$ is defined in Eq.~\ref{eq:likelihood}. For a good fit, $\chi^2/DOF \sim 1$, where $DOF$ is the number of data points minus the total number of free parameters. Although this calculation of DOF assumes that the model is linear as a function of the free parameters~\cite{Melchior}, this reduced $\chi^2$ provides a useful rule of thumb to check if the fit is good. The $\chi^2$/DOF are shown in Table~\ref{tab:results}. We can see that none of the three models can adequately fit the data since $\chi^2$/DOF $\sim 10$ for all the models. This is in accord with the results in ~\cite{Ganguly}, who also found that none of the three models provide a good fit to the spectral lag data for GRB 1606025B.  However,~\cite{Du} had found that all the LIV models provide a good fit to the spectral lag data of GRB  190114C. Therefore, our results show when we stack the   spectral lag  data from different GRBs, no one model provides a robust description of the time lags.

\begin{figure*}
    \centering
    \includegraphics[width=14cm,height=14cm]{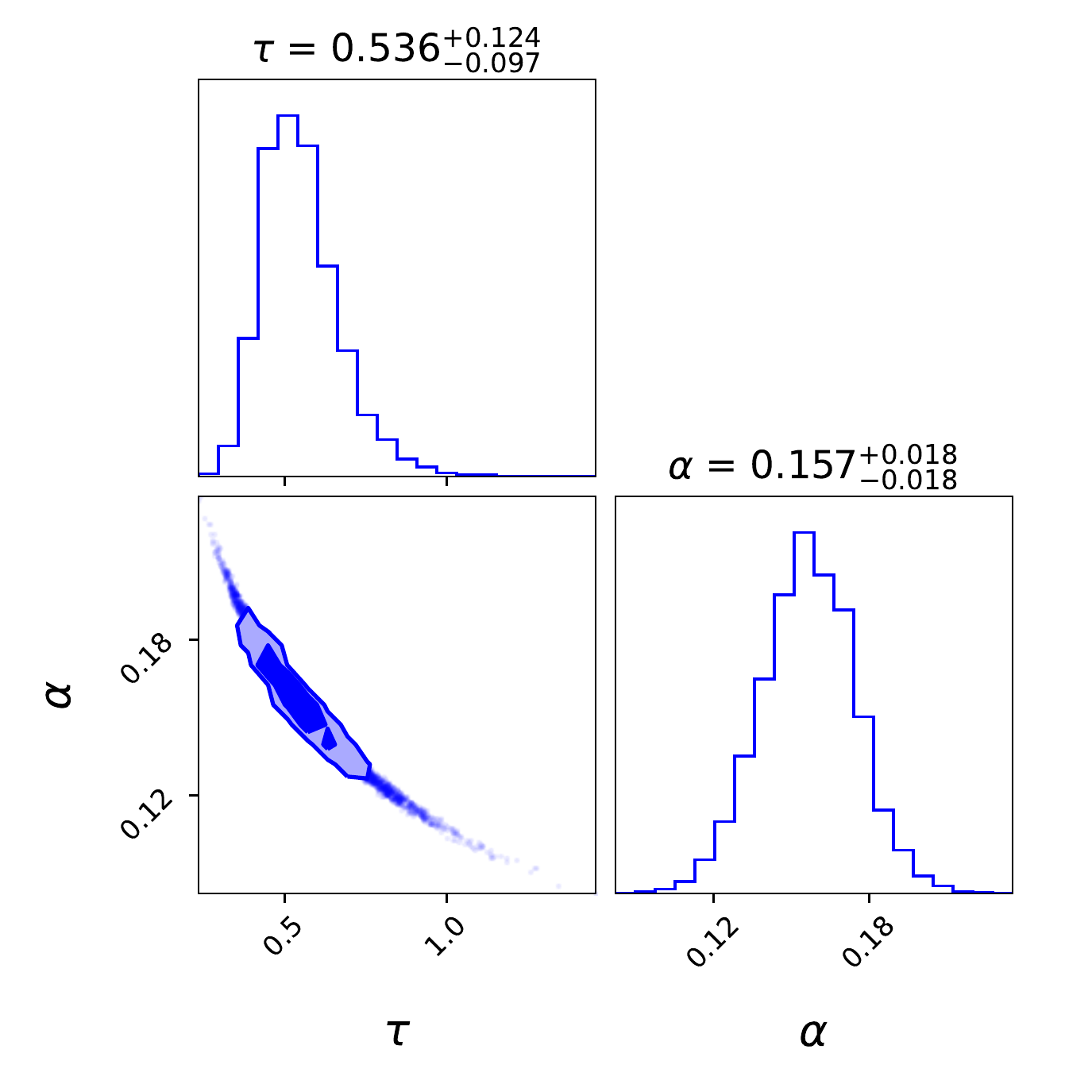}
    \caption{The marginalized 68\% and 90\% credible regions for the parameters of the null hypothesis of the only intrinsic astrophysical emission model (cf. Eq.~\ref{eq:delt_int}).  The marginalized best-fit estimates for $\tau$ and $\alpha$ are shown in the figure.}
    \label{fig:null}
\end{figure*}

\begin{figure*}
    \centering
    \includegraphics[width=15cm,height=15cm]{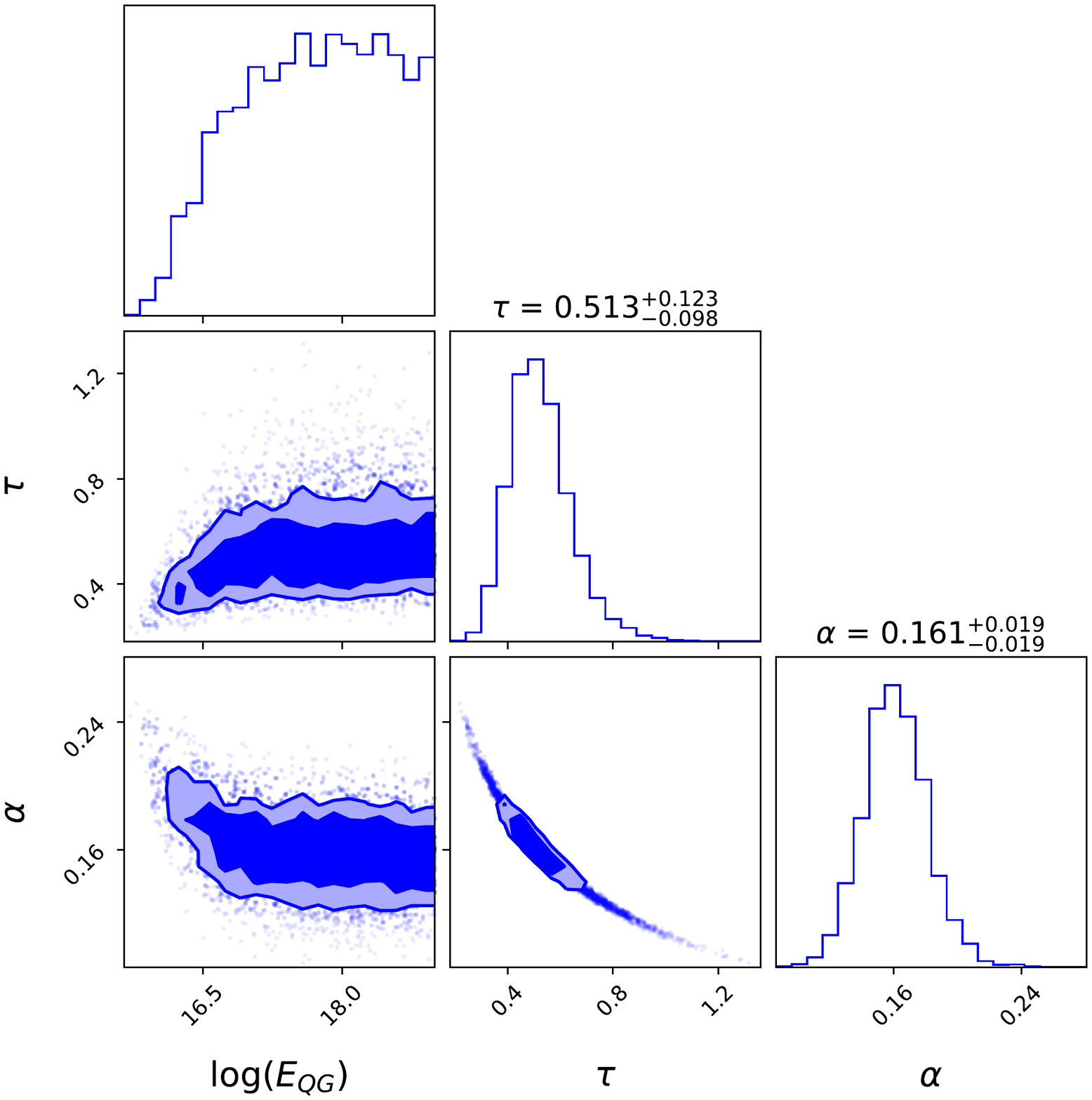}
    \caption{The marginalized 68\% and 90\% credible regions for the linear LIV model, corresponding to $n=1$   in Eq.~\ref{eq:deltaliv}. We do not get a closed contour for $E_{QG}$. Therefore, marginalized central estimates for $E_{QG}$ cannot be defined. The marginalized best-fit estimates for $\tau$ and $\alpha$ can be found in the figure.}
    \label{fig:f1}
\end{figure*}

\begin{figure*}
    \centering
    \includegraphics[width=14cm,height=14cm]{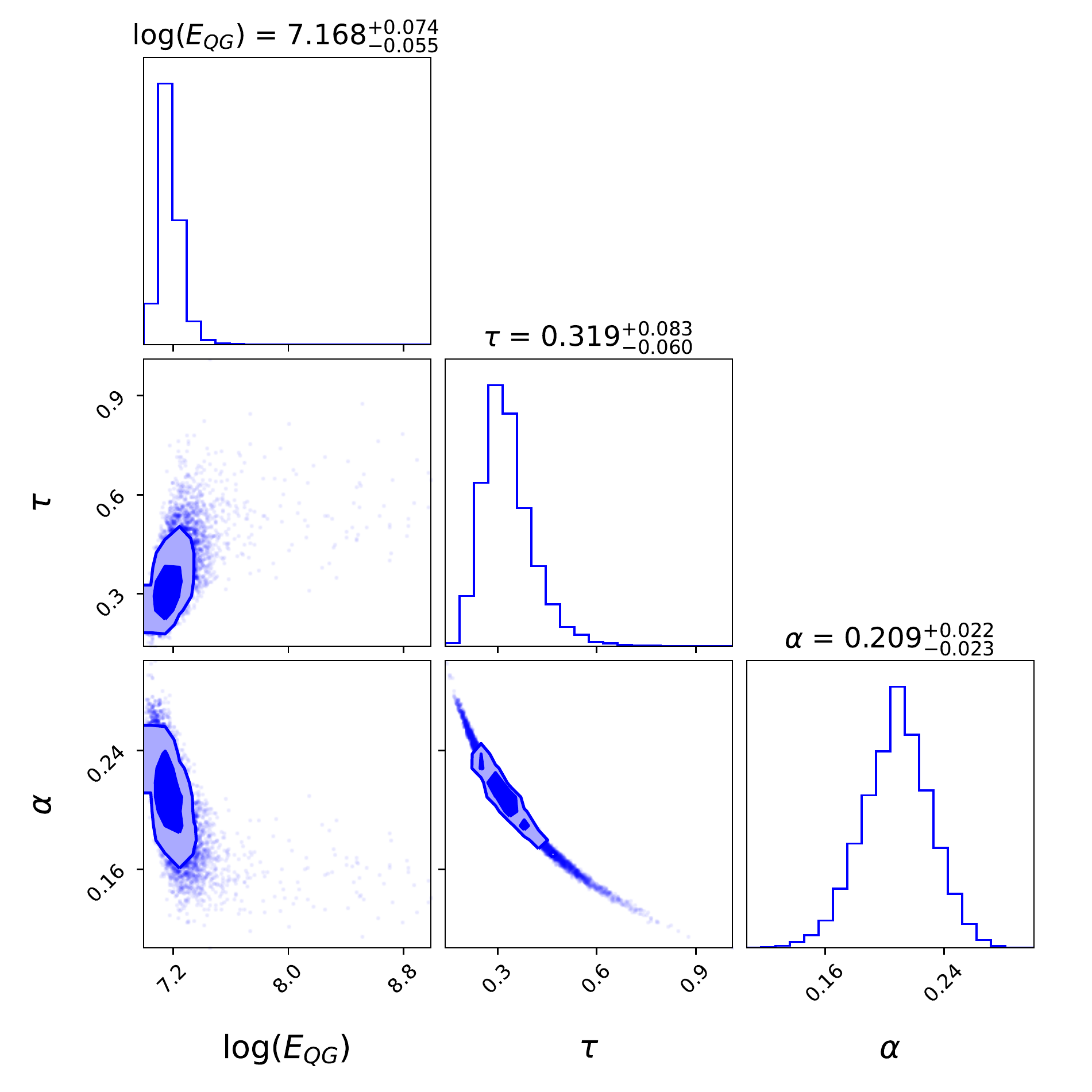}
    \caption{The marginalized 68\% and 90\% credible regions for the quadratic LIV parameters, corresponding  $n=2$ defined in Eq.~\ref{eq:deltaliv}.  }
    \label{fig:f2}
\end{figure*}

 The corresponding 68\% and 90\% marginalized credible intervals for all the free parameters can be found  in Figs.~\ref{fig:null}, ~\ref{fig:f1}, and ~\ref{fig:f2} for the null hypothesis,  $n=1$ LIV, and $n=2$ LIV models, respectively. Unlike ~\cite{Wei,Pan,Du}, we do not get closed contours for $E_{QG}$ (with $E_{QG}$ less than Planck scale) for the linear LIV model. 
 Consequently, we cannot obtain bound 1$\sigma$ marginalized point estimates for $E_{QG}$ for the linear model. We get closed contours only  for the quadratic LIV model, with the marginalized central estimate for $E_{QG} = (7.17^{+0.07}_{-0.055}) \times 10^{7}$ GeV. The marginalized central estimates  for all the other parameters can be found in Table~\ref{tab}.

Therefore, for the linear LIV model, we can only calculate lower limit for $E_{QG}$. For this purpose, we use the same method as in ~\cite{Ellis}, which we briefly describe. We calculate the marginal likelihood ($L_{marg}$) over the nuisance parameters ($\tau$ and $\alpha$), and  obtain the 68\% lower limit on $E_{QG}$ by  solving the following equation
\begin{equation}
    \frac{\int \limits_{E_{QG}}^{E_{\infty}} L_{marg}(x) dx}{\int \limits_{E_{0}}^{E_{\infty}} L_{marg}(x) dx} =0.68, 
    \label{eq:lowerlimit}
\end{equation}
where similar to ~\cite{Ellis}, $E_{\infty}$ indicates the maximum value used for fixing the normalization and is chosen to be the Planck scale equal to $10^{19}$ GeV. $E_{0}$ is the lower limit used in calculating the integral is equal to $10^6$ GeV, which is the lower limit of our prior on $E_{QG}$ (cf. Table~\ref{priortable}). This lower limit on  $E_{QG}$ for the linear LIV  model can be found in Table~\ref{tab}.

\begin{table}
    \centering
    
    \begin{tabular}{|c|c|c|c|}
    
    \hline
    \textbf{Parameter}&{\textbf{Prior}} & {\textbf{Minimum}} &  {\textbf{Maximum}}\\
        \hline 
    $\alpha$  & Uniform & -1 & 1  \\
    $\tau$ & Uniform & -10 & 10  \\
    $\log_{10} (E_{QG}/GeV)$  & Uniform & 6 & 19  \\
     \hline 
 \end{tabular}
 \caption{\label{priortable} List of priors used for the parameters of the three models (cf. Eq.~\ref{eq:delta t}). Note that $E_{QG}$ is not used for the null hypothesis of only intrinsic emission.}
\end{table}

To calculate the Bayesian evidence, we again use the same priors described in Table~\ref{priortable}.  The evidence was computed using the  {\tt Dynesty}~\cite{dynesty}  package in Python, which is based on the Nested sampling algorithm~\cite{nestedsampling,Buchner}.  As discussed in Sec.~\ref{sec:modelcomp}, the  Bayes factor  is the ratio of the Bayesian Evidence. These Bayes factors are shown in Table~\ref{tab:results}. We can see that the Bayes factor for the $n=1$ LIV hypothesis is close to  one. According to Jeffrey's scale~\cite{Trotta}, this corresponds to inconclusive evidence for any of the linear LIV model.  For the quadratic LIV model, we get a Bayes factor of about 25. According to the Jeffreys scale, this only corresponds to strong evidence and not decisive evidence.
Therefore, we conclude that when we stack the spectral lag data for GRB 160625B, GRB 190114C, and the 35 GRB sample analyzed in ~\cite{Ellis}, Bayesian model comparison does not show a decisive evidence  for either the linear or quadratic LIV over pure astrophysical emission.

\begin{table}
\begin{center}
\begin{tabular}{|c |  c |   c |   c| }
\hline
&  \textbf{Intrinsic} & \textbf{ (n=1) LIV} & \textbf{(n=2) LIV}  \\
\hline
\textbf{$\alpha$}  &$0.16 \pm 0.02$ & $0.16 \pm 0.02$ & $0.16 \pm 0.02$ \\
\textbf{$\tau$ (sec)} &$0.54 \pm 0.1 $ &$0.52 \pm 0.1$ & $0.54 \pm 0.1$  \\
\textbf{$\log_{10} (E_{qg})$ (GeV)} &  &$>16.079$ (68\% c.l.) &  $ 7.17^{+0.07}_{-0.055} $   \\
\hline
\end{tabular}
\caption{\label{tab} Best-fit values of the models for the three hypotheses considered. The intrinsic hypothesis is given by Eq.~\ref{eq:nullhypothesis} and the two LIV models are obtained by plugging $n=1,2$ in the second term  in Eq.~\ref{eq:delta t}. For the linear LIV model, we only report 68\% c.l. lower limits on $E_{QG}$.}
\end{center}
\end{table}

\begin{table}
\begin{center}
\begin{tabular}{|c| c | c | c|}
\hline
&  \textbf{No LIV} & \textbf{ (n=1) LIV} & \textbf{(n=2) LIV}  \\
\hline
\(\chi^2/\rm{DOF} \) & 970/89 & 969/88 & 936/88 \\

Bayes Factor & -  & 0.155 &  25\\ \hline
\end{tabular}
\caption{\label{tab:results} Bayesian statistical significance of Lorentz invariance violation (LIV)   for the two models (linear and quadratic LIV) as compared to the null hypothesis of only intrinsic emission. We also provide the $\chi^2$/DOF for all the three models. We can see that none of the three models provide a good fit to the spectral lag data since $\chi^2/DOF \sim \mathcal{O}(10)$. The Bayes factor shows negligible evidence for the linear LIV model and strong evidence for the quadratic LIV model.}
\end{center}
\end{table}


\section{Analysis using a  constant intrinsic lag}
\label{sec:const}
One issue  with the previous analysis is that the $\chi^2$/DOF for all the three models is greater than one. This shows that the parametric form  used to characterize the  astrophysical emission cannot self-consistently describe all the GRBs. Although, other models such as the magnetic jet model have been invoked to model the intrinsic emission~\cite{Chang12}, the intrinsic emission mechanism of GRBs is still unknown. There might be other parametric forms for $\Delta t_{int}$ or different terms for different GRBs. Although, the expression used in Eq.~\ref{eq:delt_int} seems reasonable, given the  correlation between the spectral lags and energy observed in GRB 1606025B~\cite{Wei}  and GRB 190114C~\cite{Du} at low energies, there could be other parametric forms for $\Delta t_{int}$ or different values for $\tau$ and $\alpha$  (in Eq.~\ref{eq:delt_int}) for different GRBs.

Therefore, to circumvent any systematic effects due to the aforementioned intrinsic spectral lag, we redo our search for LIV using the same three datasets by assuming a constant intrinsic lag similar to ~\cite{Ellis,Shao,Pan15,Xu1,Xu2,Liu,Wei17,Zou}, along with  adding an unknown intrinsic scatter to the observational errors in our likelihood. The addition of an unknown intrinsic scatter is common in Cosmology (eg.,~\cite{Gopika,Pradyumna,Bora}) and has also previously been used in LIV searches~\cite{Wei17}.
The observed spectral lag can then be written as:
\begin{equation}
  \Delta t = b(1+z)   -\left(\frac{1+n}{2H_0}\right)\left(\frac{E^n - E_0^n}{E_{QG,n}^n}\right) K(z),
  \label{eq:const}
\end{equation}
where $K(z)$ is given in Eq.~\ref{eq:Kz}, $b$ is the unknown constant spectral lag, and all other terms are same as in Eq.~\ref{eq:deltaliv}. We also add an unknown intrinsic scatter in quadrature to the  total error, defined earlier in Eq.~\ref{eq:totalerror}. Our results for the Bayesian model comparison with these assumptions and using the same procedure as in Sec.~\ref{sec:results} can be found in Table~\ref{tab:const}. As we can see, the Bayes factors for both the LIV models are  close to one, indicating that there is no evidence for LIV, when we assume a constant intrinsic lag and include an unknown systematic error.

\begin{table}
\begin{center}
\begin{tabular}{|c| c | c | c|}
\hline
&  \textbf{No LIV} & \textbf{ (n=1) LIV} & \textbf{(n=2) LIV}    \\ \hline
Bayes Factor & -  & 0.2 &  0.8\\ \hline
\end{tabular}
\caption{\label{tab:const} Bayesian evidence for LIV searches using the same datasets as  in Table~\ref{tab:results}, but  assuming a constant intrinsic lag (cf. Eq.~\ref{eq:const}), along with an  unknown intrinsic scatter.   }
\end{center}
\end{table}

\section{Analysis with  different intrinsic parameters per GRB sample}
\label{sec:diff}
Even though using a unified model for the intrinsic emission seems reasonable, if all GRBs can be described by the same Physics,  since the time durations of GRBs  span over six orders of magnitude, it is unlikely that different GRBs  have the same intrinsic time lag between two fixed energy bands. Therefore,  the intrinsic emission mechanism of  GRBs  described in Eq.~\ref{eq:delt_int} could have GRB dependent values for $\tau$ and $\alpha$. In ~\cite{Pan}, same values for $\tau$ and $\alpha$ were used to analyze the spectral lag data for GRB 1606025B as well the sample in ~\cite{Ellis}. However, ~\cite{Du} used the opposite sign for $\alpha$ and $\tau$ compared to ~\cite{Pan} or earlier works~\cite{Wei,Ganguly}. Taking a cue from this, we assume that the intrinsic emission mechanism for GRB 190114C can be fit with parameters $\alpha_1$ and $\tau_1$, whereas for all other GRBs, the best-fit intrinsic parameters are given by  $\alpha_2$ and $\tau_2$. We now redo the same  earlier analysis with this assumption.

\begin{figure*}
    \centering
    \includegraphics[width=17cm,height=17cm]{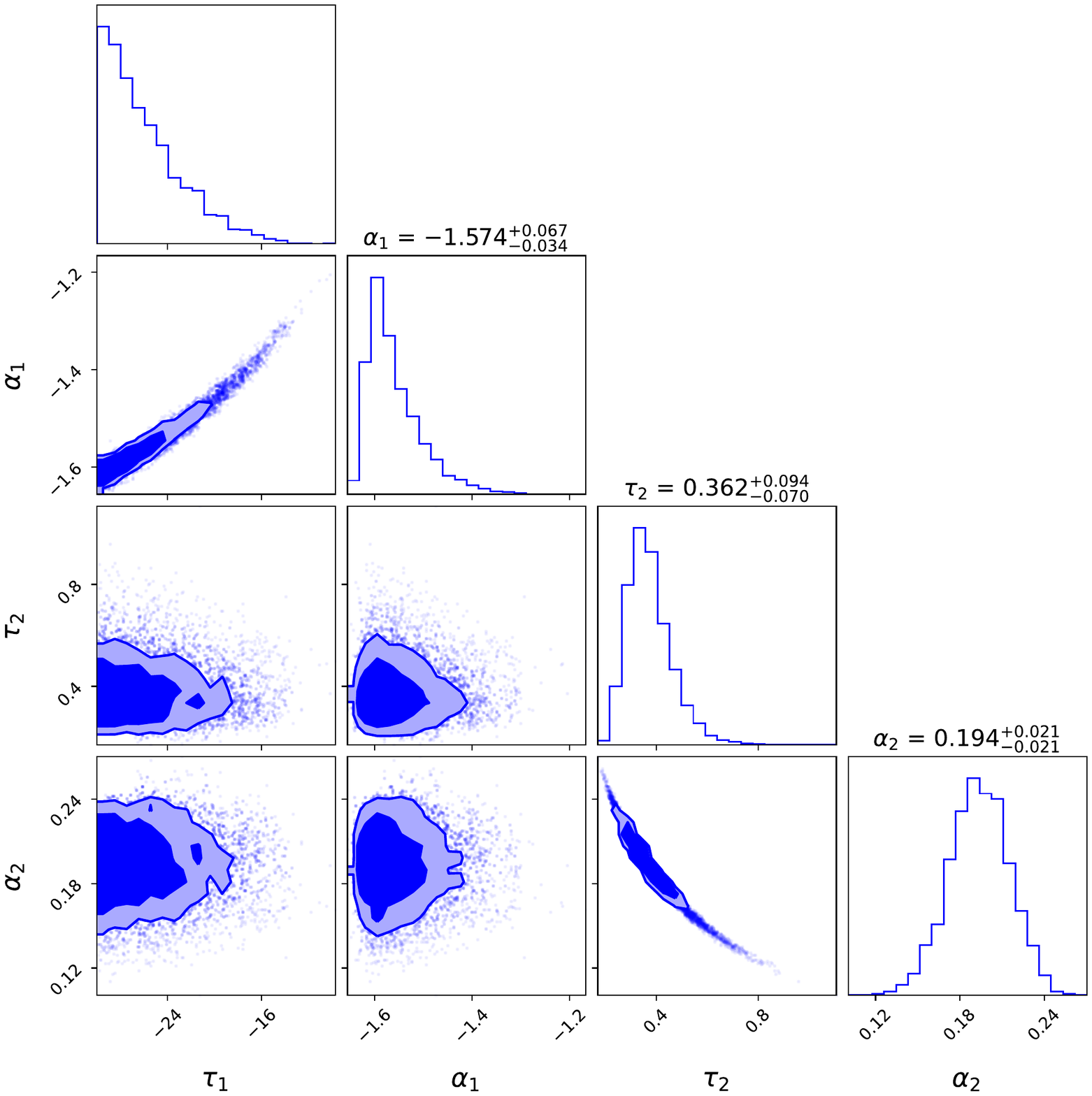}
    \caption{The marginalized 68\% and 90\% credible regions for the parameters of the null hypothesis (cf. Eq.~\ref{eq:delt_int}) after using the intrinsic parameters $\alpha_1$, $\tau_1$ for GRB 190114C  and $\alpha_2$, $\tau_2$ for the rest. Note that we do not get closed contours for $\tau_1$.}
    \label{fig:null_diff}
\end{figure*}

\begin{figure*}
    \centering
    \includegraphics[width=17cm,height=17cm]{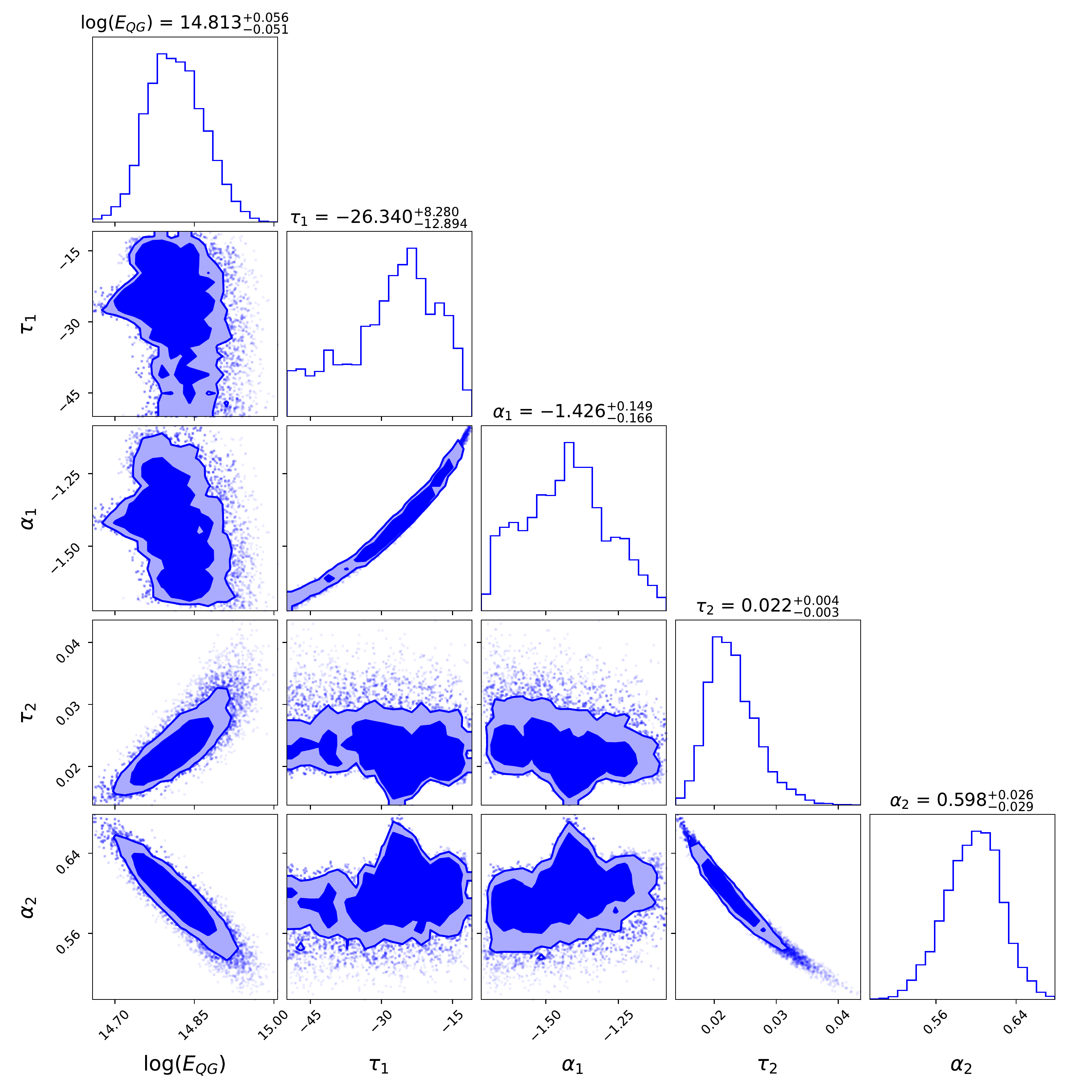}
    \caption{The marginalized 68\% and 90\% credible regions for the linear LIV model  in Eq.~\ref{eq:deltaliv} after using different intrinsic parameters for GRB 190114C versus the rest, similar to that discussed in the caption of Fig.~\ref{fig:null_diff}.}
    \label{fig:f1_diff}
\end{figure*}

\begin{figure*}
    \centering
    \includegraphics[width=20cm,height=17cm]{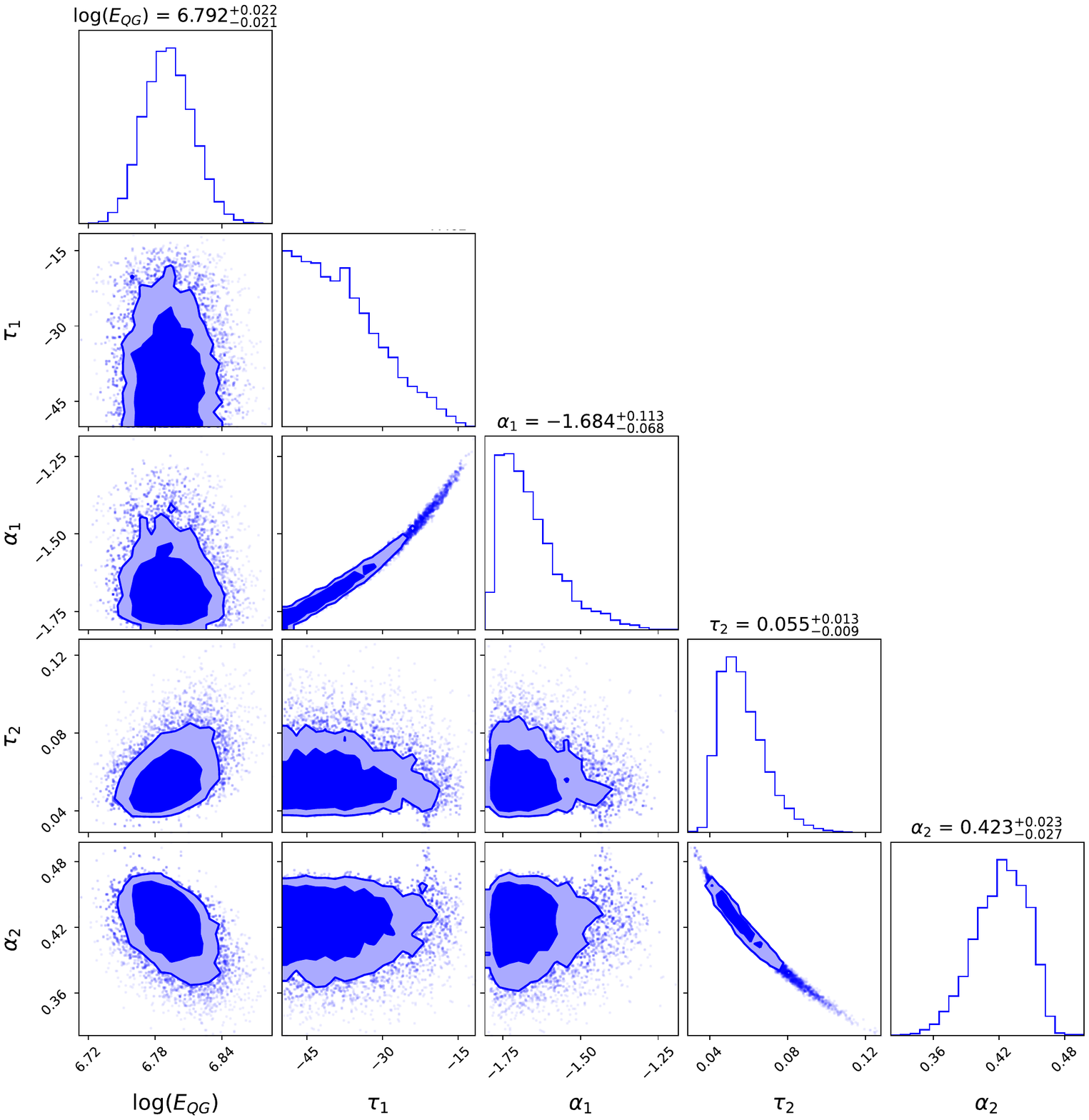}
    \caption{The marginalized 68\% and 90\% credible regions for the quadratic LIV parameters, corresponding  $n=2$ after using different intrinsic parameters for GRB 190114C versus the rest, as discussed in the caption of Fig.~\ref{fig:null_diff}.}
    \label{fig:f2_diff}
\end{figure*}

The marginalized credible intervals for the all parameters of the null hypothesis and the two LIV hypotheses with these sets of assumptions can be found in Fig.~\ref{fig:null_diff}, Fig. ~\ref{fig:f1_diff},Fig.~\ref{fig:f2_diff}, respectively. We find that we do not get closed contours for $\tau_1$ for  both the null and quadratic LIV model. Another major difference, compared to the quadratic model analyzed in Sec.~\ref{sec:results} is  that now we get  closed marginalized 68\% intervals for $E_{QG}$ for the linear LIV model, unlike in the previous case when we used the same intrinsic parameters for all the GRBs The best-fit values for all the parameters can be found in Table~\ref{tab_diffalpha}. The $\chi^2$/dof along with the Bayes factor for both the LIV hypotheses are shown in Table~\ref{tab:results_diffalpha}. The Bayes factors for both the LIV models  is $\mathcal{O}(10^{77})$, indicating decisive evidence for LIV.  Therefore, upon  relaxing the assumption of the same intrinsic parameters elevates the Bayes factors to over 70 orders of magnitude.These Bayes factors are of the same order of magnitude as those obtained using only GRB 1901114C~\cite{Du}. The $chi^2$/dof for both the LIV hypotheses are also smaller than those in Table~\ref{tab:results}, although they are still greater than one.

\begin{table}
\begin{center}
\begin{tabular}{|c |  c |   c |   c| }
\hline
&  \textbf{Intrinsic} & \textbf{ (n=1) LIV} & \textbf{(n=2) LIV}  \\
\hline
\textbf{$\alpha_1$}  &$-1.57^{+0.07}_{-0.03}$ & $-1.43^{+0.15}_{-0.17}$ & $-1.68^{+0.01}_{-0.07}$ \\
\textbf{$\tau_1$ (sec)} & -  &$26.34^{+8.2}_{-12.9}$ & -  \\
\textbf{$\alpha_2$}  &$0.19 \pm 0.02$ & $0.6 \pm 0.03$ & $0.42^{+0.02}_{-0.03}$ \\
\textbf{$\tau_2$ (sec)} & $0.36^{+0.09}_{-0.07}$ &$0.022^{+0.004}_{-0.003}$ & $0.055 \pm 0.01$  \\
\textbf{$\log_{10} (E_{qg}$ (GeV))}   & - &$14.81 \pm 0.05$ &  $ 6.79 \pm  0.02 $   \\
\hline
\end{tabular}
\caption{\label{tab_diffalpha} Best-fit values of the parameters for the three hypotheses considered, when using different intrinsic parameters $\tau_1$, $\alpha_1$ (cf. Eq.~\ref{eq:delt_int})  for GRB 190114C as compared to the rest, described by $\tau_2$ and $\alpha_2$. We could not get closed bound intervals for $\tau_1$ for null and linear LIV models and hence those values are not reported.}
\end{center}
\end{table}

\begin{table}
\begin{center}
\begin{tabular}{|c| c | c | c|}
\hline
&  \textbf{No LIV} & \textbf{ (n=1) LIV} & \textbf{(n=2) LIV}  \\
\hline
\(\chi^2/\rm{DOF} \) & 318604/87 & 402/86 & 595/86 \\
Bayes Factor & -  & $1.8 \times 10^{78}$ &  $6.9 \times 10^{77}$\\ \hline
\end{tabular}
\caption{\label{tab:results_diffalpha} Bayesian statistical significance of LIV searches using different intrinsic parameters for GRB 190114C as compared to the rest. The Bayes factor which we obtain is of the same order of magnitude as that in ~\cite{Du} and point to decisive evidence for both the LIV models. The first row shows the $\chi^2$/dof for all the three hypotheses. For the LIV hypotheses, these are smaller than those in  Table~\ref{tab:results}, although still greater than one.}
\end{center}
\end{table}

\section{Analysis with rest-frame spectral lags}
\label{sec:sourceframe}
All the  data used for LIV searches presented so far consist of spectral lags between two fixed energy bins in the observer frame. Since this data contains GRBs with different redshifts, the corresponding energy band in the source frame would be  different. This could introduce an energy-dependent spectral lag and/or cause an additional systematic uncertainty in the constraints on LIV~\cite{Wei17,Wei2021}. Ukwatta et al~\cite{Ukwatta2012} showed that there is a large scatter in the correlation between the observer-frame lags and the source-frame lags for the same GRB sample, implying that the observer- frame lag does not faithfully represent the rest-frame lag. In other words, the observer-frame lags would be strongly biased, as  they  record different pairs of the intrinsic light curves. To ameliorate this, similar to ~\cite{Wei17}, we carry out an independent analysis for LIV search by choosing two fixed energy bands in the source frame, and thereby obtaining the observed lag  in the observer frame  by the relation $E_{obs}= E_{source}/(1+z)$. Such an analysis would eliminate any potential biases due to energy-dependent effects.

To carry out a search for LIV using the source frame spectral lags, we use the dataset compiled in ~\cite{Bernardini}, which were also used in ~\cite{Wei17}. This sample consists of rest-frame lags of 56 GRBs measured using the Neil Gehrels SWIFT observatory, with redshifts ranging from 0.35 to 5.47. These lags were measured between fixed source frame energy bands of 100-150 keV and 200-250 keV. These lags have been tabulated in ~\cite{Wei17,Ukwatta2012}, and we use them for our analysis.

To search for LIV using these rest frame lags, we fit the observer-frame spectral lag ($\Delta (t)$) to a  sum of intrinsic lag and LIV-induced lag, which can be written  similar to Eqs~\ref{eq:delt_int} and ~\ref{eq:deltaliv}:

\begin{equation}
\Delta t  = \tau (1+z) \Big[ \Big(\frac{E}{keV}\Big)^{\alpha}-\Big(\frac{E_0}{keV}\Big)^{\alpha}\Big] -\frac{1+n}{2H_0 (1+z)^n}\left(\frac{E^n - E_0^n}{E_{QG,n}^n}\right) K(z)
\label{eq:sourceframe}
\end{equation}
In Eq.~\ref{eq:sourceframe}, the first term represents the intrinsic lag and the second term represents the lag due to LIV. Note that unlike Eq.~\ref{eq:delt_int} and Eq.~\ref{eq:deltaliv}, $E$ and $E_0$ in Eq.~\ref{eq:sourceframe} now denote the upper and lower energy intervals in the {\em source frame}, respectively. For our dataset in ~\cite{Bernardini}, $E$ and $E_0$ correspond to 125 keV and 225 keV, respectively. Note that $K(z)$ in Eq.~\ref{eq:sourceframe} is same as that in Eq.~\ref{eq:Kz}. 

We point out  that this analysis differs from ~\cite{Wei17} in a couple of aspects. We have fit this data  to a LIV model assuming $s_{\pm}=+1$, wheres the fit in  ~\cite{Wei17} was done  assuming $s_{\pm}=-1$. Secondly, instead of fitting for a constant intrinsic emission (as done in ~\cite{Wei17}), we have used a model, where  the intrinsic spectral lag is proportional to the source-frame energy. This is similar to the model used for fitting observer frame intrinsic spectral lags (cf. Eq.~\ref{eq:delt_int}), except that the rest-frame energies have been used in Eq.~\ref{eq:sourceframe}.

We now carry out Bayesian model comparison in the same way as in Sect.~\ref{sec:results}, to see if the rest frame spectral lags show statistically significant evidence for LIV Our results are shown in Table~\ref{tab:restframeresults}. We find $\chi^2$/DOF is greater than 1 for all the three models, indicating that we don't see a good fit. This is also in accord with the results in ~\cite{Wei17}, who added an intrinsic scatter to the observed errors. When we do Bayesian model comparison, we find that only the linear LIV model shows marginal evidence, whereas the quadratic LIV model shows negligible evidence. Therefore, none of the two models show decisive evidence for Lorentz violation using the dataset compiled in ~\cite{Bernardini}. For a more definitive test one would need more statistics on the rest frame spectral lags.

\begin{table}
\begin{center}
\begin{tabular}{|c| c | c | c|}
\hline
&  \textbf{No LIV} & \textbf{ (n=1) LIV} & \textbf{(n=2) LIV}  \\
\hline
\(\chi^2/\rm{DOF} \) & 77/54 & 613/53 & 259/53 \\

Bayes Factor & -  & 9.6 &  0.8\\ \hline
\end{tabular}
\caption{\label{tab:restframeresults} Bayesian statistical significance of LIV searches using rest-frame spectral lag data compiled in ~\cite{Bernardini}, done in the same way as in Table~\ref{tab:results}. The Bayes factors for the linear and quadratic LIV model indicate  evidence and no evidence, respectively.}
\end{center}
\end{table}

\section{Constraints on LIV Standard Model Extension}
\label{sec:SME}
We now use  stacked searches for LIV to constrain the parameters of the LIV Standard Model Extension (SME)~\cite{Kostelecky}, and also calculate its statistical significance as compared to the null hypothesis using Bayesian model comparison techniques.
The corresponding expression for $\Delta t_{LIV}$ in the  SME models is given by ~\cite{Mewes1,Mewes2,Wei2,Du}:
\begin{equation}
\Delta_{LIV} \approx -(d-3) \left(E^{d-3} - E_0^{d-4}\right) \times \int \limits_{0}^{z} \frac{(1+z')^{d-4}}{H_{z'}}dz' \sum_{jm} {}_0 Y_{jm}(\hat{n})c_{(I)jm}^{(d)},
\label{eq:SME}
\end{equation}
where $d$ represents the mass dimension of the Lorentz-violating operator in the Lagrange density of the SME.  In Eq~\ref{eq:SME}, the Lorentz violating term of the SME is encapsulated in the sum of the spin weighted spherical harmonic coefficient $\sum_{jm} {}_0 Y_{jm}(\hat{n})c_{(I)jm}^{(d)}$, where $\hat{n}$ is a unit vector pointing to the source and $jm$ contain the eigenvalues of the total angular momentum for each combination of the terms. The spherical polar coordinates ($\theta$,$\phi$) associated with $\hat{n}$ are defined with respect to a Sun-centered frame with $\theta=90 -\delta$ and $\phi=RA$~\cite{Kostelecky02} for a source at right ascension = $RA$ and declination =$\delta$.   Similar to ~\cite{Du}, we assume that this term is positive, so that the high energy photons arrive later compared to the low energy ones, implying a negative spectral lag in the presence of LIV.

\begin{figure*}
    \centering
    \includegraphics[width=17cm,height=17cm]{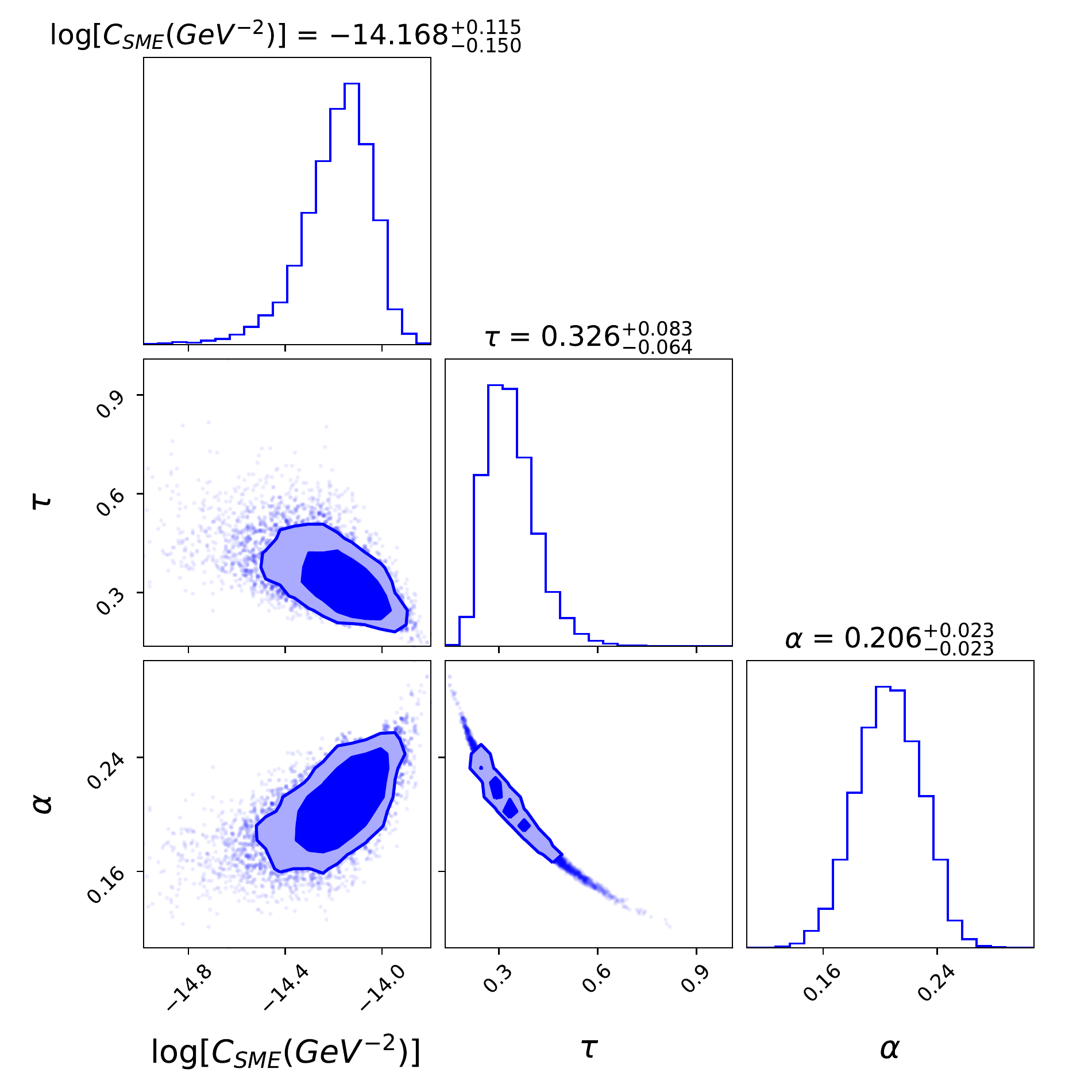}
    \caption{The marginalized 68\% and 90\% credible regions for the parameters of $d=6$ LIV SME (cf. Eq.~\ref{eq:SME}). The quantity $C_{SME}$ is equal to $\sum_{jm} {}_0 Y_{jm}(\hat{n})c_{(I)jm}^{(6)}$  }
    \label{fig:SME_d6}
\end{figure*}

\begin{figure*}
    \centering
    \includegraphics[width=17cm,height=17cm]{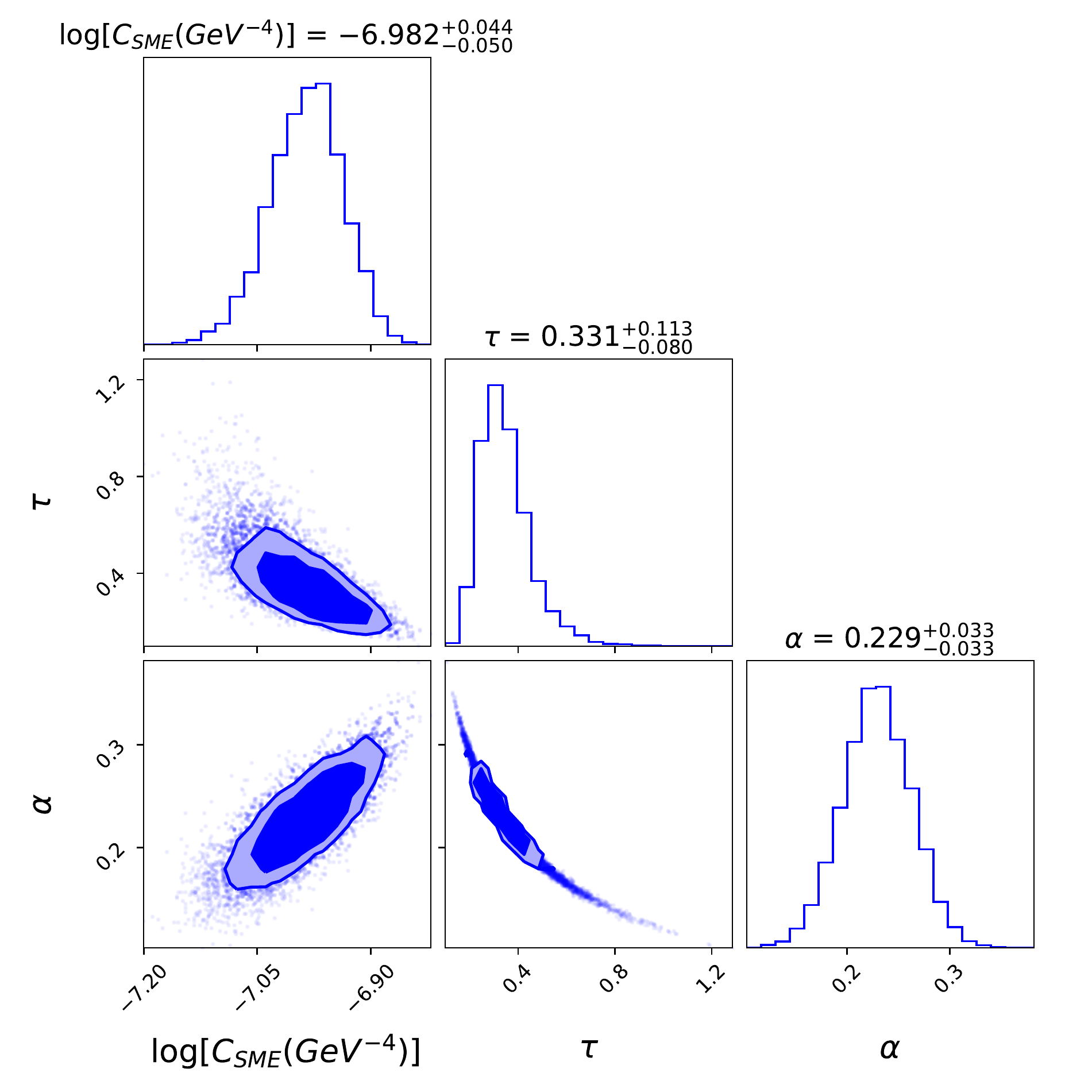}
    \caption{The marginalized 68\% and 90\% credible regions for the parameters of $d=8$ LIV SME (cf. Eq.~\ref{eq:SME}). The quantity $C_{SME}$ is equal to $\sum_{jm} {}_0 Y_{jm}(\hat{n})c_{(I)jm}^{(8)}$  }
    \label{fig:SME_d8}
\end{figure*}

We now fit the observed spectral lag data for all the 37 GRBs to the sum of LIV and intrinsic emission (cf. Eq.~\ref{eq:delta t}), where $\Delta_{LIV}$ is given by Eq.~\ref{eq:SME} and the intrinsic model given  by Eq.~\ref{eq:delt_int}. We perform our fits for two values of $d$, namely $d=6$ and $d=8$. The 68\% and 90\% allowed regions for the SME coefficients for $d=6$ and $d=8$ can be found in Fig.~\ref{fig:SME_d6} and Fig.~\ref{fig:SME_d8}, respectively. For brevity, the spin-weighted spherical harmonic coefficient 
$\sum_{jm} {}_0 Y_{jm}(\hat{n})c_{(I)jm}^{(d)}$ have been denoted  by $C_{SME}$ in these plots.
We find that we can get closed contours for both $d=6$ and $d=8$. These best-fit values are summarized in Table~\ref{tab:results_SME}.
Finally, the results from Bayesian model comparison for both these hypotheses are shown in Table~\ref{tab:modelcomp_SME}. As we can see, the $d=6$ and $d=8$ SME show very strong and decisive  evidence, respectively compared to the null hypothesis of only intrinsic  emission. However as before, the $\chi^2$ is still greater than one for both these models.

\begin{table}
\begin{center}
\begin{tabular}{|c |c |    c | }
\hline
& \textbf{ ($d=6$) SME} & \textbf{($d=8$) SME}  \\
\hline
\textbf{$\alpha$}  & $0.21 \pm 0.02$ & $0.23 \pm 0.03$ \\
\textbf{$\tau$ (sec)} &  $0.33^{+0.083}_{-0.064}$  
 & $0.33^{+0.11}_{-0.08}$   \\
\textbf{
$\sum_{jm} {}_0 Y_{jm}(\hat{n})c_{(I)jm}^{(d)}$}    & $10^{(-14.2\pm 0.1)} (\text{GeV}^{-2})$  &  $ 10^{(-7.0^{+0.04}_{-0.05})}  (\text{GeV}^{-4})$   \\
\hline
\end{tabular}
\caption{\label{tab:results_SME} Best-fit values   for the LIV SME coefficient  $\sum_{jm} {}_0 Y_{jm}(\hat{n})c_{(I)jm}^{(d)}$  for $d=6$ and $d=8$ in Eq.~\ref{eq:SME}. }
\end{center}
\end{table}

\begin{table}
\begin{center}
\begin{tabular}{|c| c | c | c|}
\hline
&  \textbf{No LIV} & \textbf{ ($d=6$) SME} & \textbf{($d=8$) SME}  \\
\hline
\(\chi^2/\rm{DOF} \) & 969/89 & 751/88 & 936/88 \\
Bayes Factor & -  & 62 &  55300\\ \hline
\end{tabular}
\caption{\label{tab:modelcomp_SME} Bayesian statistical significance and $\chi^2$/dof of LIV SME for $d=6$ and $d=8$. Note that the null hypothesis of only intrinsic astrophysical emission is the same as that in Table~\ref{tab:results}. Both the SME hypotheses have $\chi^2$/dof  greater than 1. The $d=6$  and $d=8$  models show very strong and decisive evidence, respectively. }
\end{center}
\end{table}

\section{Conclusions}
\label{sec:conclusions}
A large number of works have analyzed the spectral lag data of individual GRBs, to look for an energy dependent speed of light, characteristic of a signature of LIV. The spectral time lags have been modelled as a sum of intrinsic time lags due to astrophysical emission along with an  energy-dependent speed of light, characteristic of Lorentz violation. These works have  found evidence for LIV with varying levels of significance~\cite{Ellis,Wei,Ganguly,Pan,Du}.
If these spectral time lags can be self-consistently described by a unified model, which includes LIV, one would expect the significance to be enhanced, when one stacks the spectral lag data from different GRBs.

Therefore, to test this {\it ansatz}, we stack the spectral lag data from GRB 190114C (19 lags)~\cite{Du}, GRB 1606025B (37 lags)~\cite{Wei} and 35 additional lags from 35 different GRBs (with one lag per GRB)~\cite{Ellis}. Therefore, in all we have a total of 91 spectral lag measurements from 37 GRBs. We then analyze the stacked data using the unified model described in Eq.~\ref{eq:delta t}. The model for the intrinsic astrophysical  contribution to the  spectral lag is described in  Eq.~\ref{eq:delt_int}. The contribution due to the energy dependent speed of light can be found in Eq.~\ref{eq:deltaliv}. We test two different LIV models corresponding to $n=1$ and $n=2$ in Eq.~\ref{eq:deltaliv}.
To evaluate this, one also needs  to model the cosmic expansion history in order to evaluate Eq.~\ref{eq:deltaliv}. For this purpose, (similar to ~\cite{Pan}), we have reconstructed  this in a  non-parametric way, using GPR with individual $H(z)$ measurements from cosmic chronometers~\cite{Haveesh}. We then carry out a Bayesian model comparison to check if the stacked data show evidence for an energy dependent speed of light caused by LIV,  over purely intrinsic astrophysical emission. The priors on the   parameters for these models used for regression and model comparison can be found in Table~\ref{priortable}.

The marginalized credible intervals for the parameters of the intrinsic only, combined intrinsic and linear LIV model, combined intrinsic and quadratic LIV model can be found in Figs.~\ref{fig:null},~\ref{fig:f1}, ~\ref{fig:f2}, respectively. For the linear LIV model, we do not obtain closed $1\sigma$ bounds. Therefore, we calculate 68\% lower limits on $E_{QG}$. This lower limit along with  the  marginalized estimates for the other parameters for all the three models can be found in Table~\ref{tab}.

As a sanity check on the viability of each of these models using the stacked data, we calculate the reduced $\chi^2$ (cf. Table~\ref{tab:results}). We find that the reduced $\chi^2$ for all the three models are $\mathcal{O} (10)$, indicating that none of the them can adequately fit the stacked data. This is contrast to the results in ~\cite{Du}, who had found that the spectral lag data of GRB 190114C show reduced $\chi^2$ of around one for all the LIV models.

Our results for Bayesian model comparison between the two LIV models and the null  hypothesis (of only intrinsic astrophysical emission) can be found in Table~\ref{tab:results}. We find that the Bayes factor for the linear LIV model is close to one, indicating that there is no preference for it. The Bayes factor for the quadratic LIV model is about 25, indicating strong evidence for the quadratic model. However, the Bayes factor does not cross 100, needed to claim decisive evidence. Therefore, we find that with the stacked data, the Bayes factor for the LIV models gets degraded compared to the decisive evidence found for GRB 190114C. 

Since the aforementioned intrinsic emission model is not a pristine fit to the data, we redid our analysis using two different assumptions for the intrinsic emission. We first assume a constant intrinsic spectral lag, but also accounted for the uncertainty by positing an unknown intrinsic scatter. The results for Bayesian model comparison with this assumptions can be found in Table~\ref{tab:const}. We find that there is no evidence for either of the two LIV hypotheses. We then used different intrinsic parameters for GRB 190114C as compared to the other GRBs in our sample. The corresponding Bayesian model comparison results can be found in Table~\ref{tab_diffalpha}, with the credible intervals for the LIV  parameters in Fig~\ref{fig:f1_diff} and Fig~\ref{fig:f2_diff}.
We find that the Bayes factors now exceed 100, indicating a decisive evidence for both the models with this assumption. However the $\chi^2$/dof is still greater than one  even with this assumption. We also provide constraints on LIV SME parameters for $d=6$ and $d=8$, which can be found in Fig.~\ref{fig:SME_d6} and Fig.~\ref{fig:SME_d8}. The Bayes factor (cf. Table~\ref{tab:modelcomp_SME})  show very strong to decisive evidence for LIV. 

Finally, we also did a search for LIV using an independent dataset consisting of only rest-frame spectral lags~\cite{Bernardini}, to obviate the fact there is could be a  large scatter in the correlation between the observer frame lags and the rest-frame lags for the same GRB~\cite{Ukwatta2012}. The model comparison results using this dataset are summarized in Table~\ref{tab:restframeresults}. The linear and quadratic models now only show marginal and no evidence for LIV, respectively. Hence, the rest frame spectral lags do not show any evidence for LIV.

Therefore, we conclude that the stacked GRB spectral lag data  do not show a smoking gun evidence for any LIV model, and  cannot be uniformly explained with the same intrinsic emission model and an energy-dependent speed of light. For all the different use cases studied, we always  get   $\chi^2$/dof much greater than one, indicating that the combined model of intrinsic and LIV cannot adequately fit the observed spectral lag data.
It is possible that one needs more complicated GRB-specific model for the intrinsic spectral lag, which depends on additional parameters besides the energy. However, with  more upcoming GRB data with spectral measurements ranging from keV to TeV, one could get better statistics with more  samples having  a spectral turnover, which could enable us to disentangle the astrophysics from possible LIV signatures. This should soon be possible with the advent of CTA, LHASSO, and other gamma-ray observatories 

\bibliography{references}
\end{document}